\pdfoutput=1
\documentclass[sigconf,9pt,nonacm]{acmart}

\usepackage{amsmath}
\usepackage{booktabs}
\usepackage{graphicx}
\usepackage{xcolor}

\setcopyright{none}
\acmConference[arXiv Preprint]{arXiv Preprint}{2026}{}
\acmYear{2026}
\acmDOI{}
\acmISBN{}
\graphicspath{{figures/}}
\title{Completion-Path Credits: Multi-Resource Control for Scale-Up Fabrics}

\author{Fan Yang}
\affiliation{%
  \institution{Institute of Computing Technology, Chinese Academy of Sciences}
  \city{Beijing}
  \country{China}}
\email{yangfan2020@ict.ac.cn}

\author{Jiaqi Liu}
\affiliation{%
  \institution{Institute of Computing Technology, Chinese Academy of Sciences}
  \city{Beijing}
  \country{China}}
\affiliation{%
  \institution{Hangzhou Institute for Advanced Study, UCAS}
  \city{Hangzhou}
  \country{China}}
\email{liujiaqi242@mails.ucas.ac.cn}

\author{Tao Jiang}
\affiliation{%
  \institution{Institute of Computing Technology, Chinese Academy of Sciences}
  \city{Beijing}
  \country{China}}
\email{jt@ncic.ac.cn}

\author{Zhan Wang}
\affiliation{%
  \institution{Institute of Computing Technology, Chinese Academy of Sciences}
  \city{Beijing}
  \country{China}}
\email{wangzhan@ncic.ac.cn}

\begin{document}

\begin{abstract}
Scale-up fabrics connecting GPUs and AI accelerators carry tensor transfers
together with remote reads, writes, atomics, and notifications over shared
target-side receiver resources.  Byte-denominated credits protect link buffers
and streaming HBM traffic, but poorly represent small operations dominated by
Atomic execution or response injection.  This paper presents
\emph{SemaCredit}, a receiver controller that admits each remote-memory
operation against a vector of target-resource demands and returns each
component when its corresponding HBM, Atomic, or response stage completes.  In
a deterministic event simulator with multipath queues, eight HBM partitions, a
serialized Atomic engine, and a response engine, SemaCredit matches a strong
per-resource byte baseline on HBM-hotspot traffic while reducing
small-operation P99 latency by 52.4\% under Atomic contention and 10.2\% under response
incast.  Application-shaped mixes show 57.7\% and 14.5\% P99 latency improvements for
AllReduce-shaped and remote-read-shaped traffic while matching byte credits on
HBM-dominated MoE traffic.
\end{abstract}


\maketitle

\section{Introduction}

Scale-up interconnects are becoming the memory-access substrate of tightly
coupled GPU and accelerator systems.  Their traffic is not a uniform stream
of bytes.  A large write primarily consumes link and HBM bandwidth; a small
Atomic may occupy a serialized execution engine for much longer than its
small request would suggest; and a read consumes both target-memory service and
reverse-path response-injection capacity.  Collective communication, expert
exchange, remote parameter access, synchronization, and completion
notification place these operations on the same endpoint.

Transport congestion control nevertheless observes the network most easily in
bytes.  Receiver-driven designs grant a bounded amount of traffic, and host
backpressure can reduce a connection's sending rate.  The Ultra Ethernet
Transport (UET) specification, for example, defines receiver congestion
control around receiver-issued credits and destination backlog
\cite{uetspec}.  MRC exposes composable host backpressure and bounded-flight
primitives \cite{mrc}.  At production scale, Meta uses receiver-driven CTS
messages when collective channel buffers become ready \cite{meta-roce}.
These mechanisms are valuable building blocks, but none by itself specifies
how a transport should divide receiver capacity among an HBM partition, an
Atomic engine, and response injection.

This distinction matters because a byte queue and a service queue answer
different questions.  A byte queue asks how much data is outstanding.  A
service queue asks how long the target will remain busy.  For streaming HBM
traffic the two are nearly equivalent.  For a fixed-cost operation they can
differ by orders of magnitude.  Furthermore, an operation may consume more
than one receiver resource in sequence, so placing it in a single receiver
queue or traffic class can move rather than eliminate head-of-line blocking.

We explore the following hypothesis: receiver admission for a
memory-semantic fabric should be denominated in the native occupancy of each
target resource, and an operation should reserve all resources on its
completion path.

SemaCredit realizes this hypothesis with a small demand vector.  For operation
$i$, the vector may contain HBM-partition occupancy, Atomic-engine service
time, and response-injection service time.  Admission succeeds only if every
required resource remains within its window.  Components are returned in
phases: Atomic debt when Atomic execution finishes, HBM debt after the memory
stage, and response debt after response injection.  This lets admission treat
a small Read request as costly when its later response will occupy the return
pipeline.

This paper makes three contributions:

\begin{itemize}
  \item It identifies a concrete mismatch between byte-denominated receiver
  control and the multi-stage service demands of scale-up remote memory
  operations.
  \item It proposes SemaCredit, a resource-selective demand-vector controller
  with phased credit return and a work-conserving arbitration rule.
  \item It evaluates SemaCredit against two strong alternatives---a single
  normalized service window and a reactive resource-exhaustion policy---and
  uses ablations and sensitivity sweeps to show when the vector is necessary
  and when byte-counted control is sufficient.
\end{itemize}

\section{Motivation and Gap}

\subsection{Motivating traffic patterns}

Scale-up AI workloads combine large tensor transfers with small
synchronization, metadata, and completion operations.  The latter carry few
bytes but often decide when the next computation or communication phase can
begin.  In an AllReduce-style phase, large tensor chunks stream to remote
memory while small reduction or completion Atomics decide when the phase can
advance.  The chunks dominate bytes, but a burst of small Atomics can still
serialize at a target Atomic engine.  In remote
parameter, KV-cache, or metadata access, a small Read request can later create
response traffic that must be injected back to the initiator before the Read
completes.  MoE dispatch gives the complementary case: skewed expert writes
primarily stress the HBM partitions that receive the tokens.  These examples
motivate a receiver controller that accounts for the work an operation creates
inside the endpoint, rather than only the bytes that arrive at the link.

\subsection{Receiver service beyond receiver bandwidth}

Prior systems have reported receiver-side bottlenecks beyond link bandwidth.
RDCA reports production measurements in which receiver memory-bandwidth
contention reduces RDMA throughput by roughly 15\% and builds queues in the
RNIC despite unused network capacity \cite{rdca}.  Meta's
distributed-training network similarly exposes GPU channel-buffer readiness to
receiver-driven CTS admission and reports a tradeoff between channel count,
buffer size, and GPU execution resources \cite{meta-roce}.  These observations
show that a network-only signal can miss the last stage of delivery.

Harmonic provides the closest evidence for operation-cost accounting: it
profiles RDMA verbs and shows that similar packet sizes can impose different
RNIC-side processing costs; for example, an ATOMIC costs roughly three times a
WRITE \cite{harmonic}.  Harmonic motivates normalized operation-cost
accounting for RNIC isolation.  Scale-up memory endpoints add independently
draining HBM partitions, Atomic engines, and response-injection paths.  The
admission question is therefore which completion-path resources a Read or
Atomic reserves and when each reservation can be released, not just what total
cost the operation carries.

Scale-up memory semantics make the last stage more heterogeneous.  Consider
three same-sized small requests.  A metadata read occupies an HBM partition
and produces a response.  An Atomic add additionally serializes at an Atomic engine.  A
notification may require visibility ordering but carry no large response.
Their ingress bytes are equal, but their service paths are not.  Conversely, a
4-KB write can consume many bytes while remaining a regular streaming HBM
operation.  Treating all four with the same byte window over-throttles the
write or under-throttles the Atomic.

Recent GPU measurements also establish that Atomic service is neither a byte
rate nor a universal constant.  Uncontended device Atomics on the measured
NVIDIA GPUs take 2.4--3.5 ns, while contended cases expose a roughly 30-ns
floor; the measured AMD devices show approximately 85--100 ns under repeated
access \cite{gpu-atomic-contention}.  We use this range to stress the
controller's sensitivity to per-opcode service estimates; an implementation
would calibrate those estimates from the target endpoint it controls.

\subsection{Why splitting contexts is still incomplete}

A tempting fix is to split traffic into more contexts, for example one credit
pool per HBM partition or one class for Atomics.  This helps when the work
mostly grows with the number of bytes: a hot HBM partition can be protected by
its own byte window.  The same trick is insufficient for semantic operations.
A small Atomic still looks cheap unless the Atomic pool is charged in
serialized service time, and a Read admitted by its target HBM pool can still
overload the response-injection path.  The missing piece is therefore a demand
model that names every completion-path resource and charges it in the
resource's native unit.

\textbf{Why one normalized cost is insufficient.}  Consider two independent
receiver resources, each with safe window $W$, and two equal-cost operations
with vectors $(W,0)$ and $(0,W)$.  A scalar window no larger than $W$ is safe
for an arbitrary workload, but cannot admit one of each operation concurrently,
even though the vector controller can.  A scalar window of $2W$ admits that
parallel pair, but also admits two operations concentrated on the first
resource and violates its safe window.  Thus no fixed scalar preserves both
per-resource bounds and work conservation for arbitrary mixes; retuning it
requires already knowing the active bottleneck.

The closest application-level mechanism is Meta's receiver-driven CTS design
\cite{meta-roce}.  A receiver sends another CTS when a collective channel
buffer is ready, and the CTS includes size and memory information.  This is a
strong precedent for receiver admission informed by endpoint state.  Its unit
of scheduling, however, is a collective chunk/channel buffer.  SemaCredit asks
whether the same idea can be generalized into a transport controller for
ordinary Read, Write, Atomic, and notification operations that traverse
multiple endpoint resources.

Receiver-driven transports such as SIRD can implement flexible receiver
policies and explicitly schedule traffic to a receiver downlink
\cite{sird}.  Homa likewise demonstrates the value of receiver-driven grants
for short messages \cite{homa}.  SemaCredit is complementary: it supplies a
resource-demand policy that such a grant mechanism could enforce after link
capacity is no longer the only receiver constraint.

\section{SemaCredit Design}

\subsection{Receiver datapath}

\begin{figure}[!t]
  \centering
  \includegraphics[width=\columnwidth]{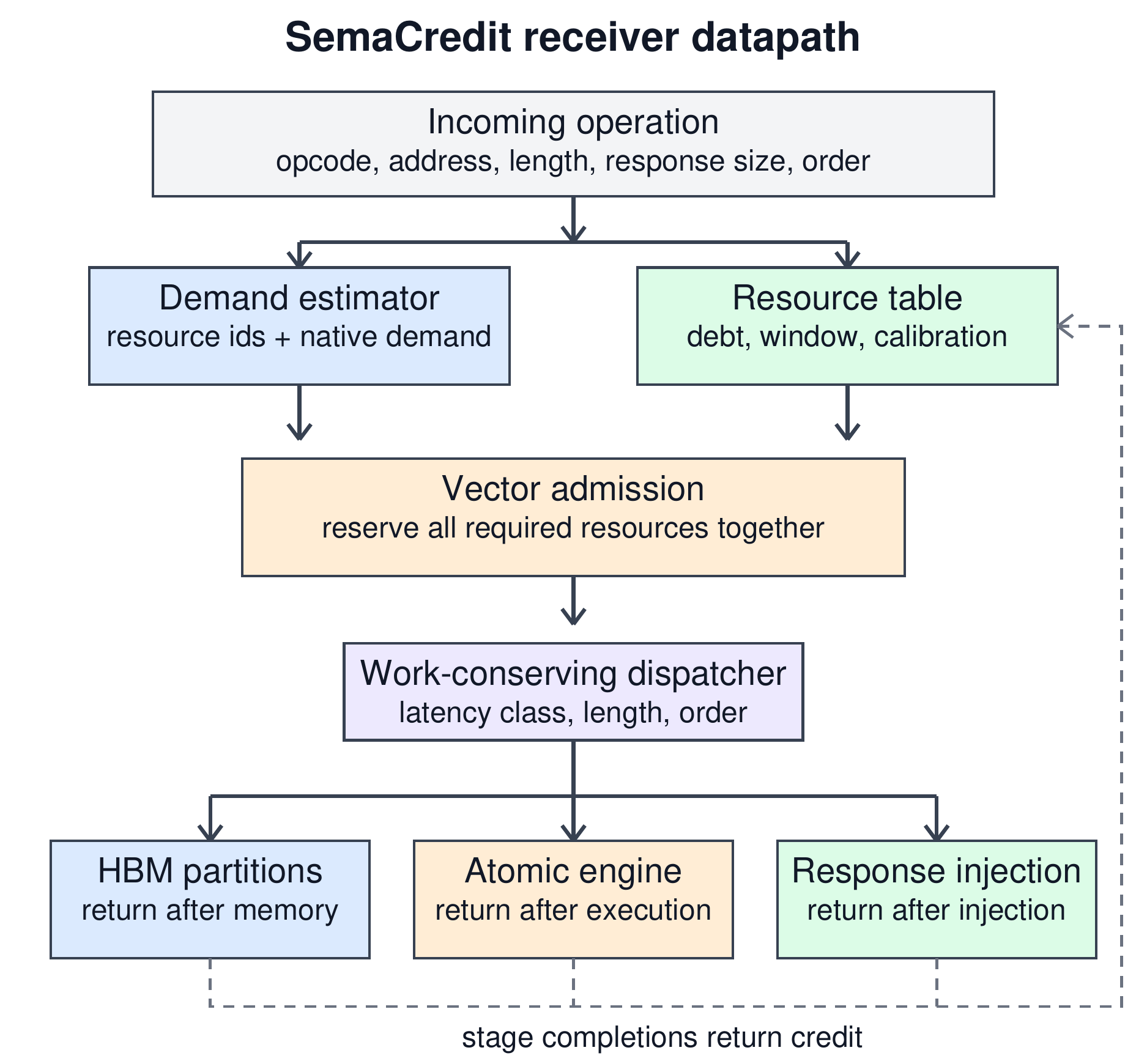}
  \caption{Logical target-side datapath.  SemaCredit estimates a demand
  vector from ordinary operation fields, reserves all required receiver
  resources before dispatch, and returns each component when the corresponding
  HBM, Atomic, or response-injection stage completes.}
  \label{fig:architecture}
\end{figure}

Figure~\ref{fig:architecture} shows the target-side organization.  An arriving
operation already carries the fields needed by an ordinary remote-memory
endpoint: opcode, address, length, response size, and ordering domain.  A
demand estimator maps those fields to an HBM partition and to any fixed-cost
Atomic or response stage.  In parallel, a small resource table supplies the
current debt, window, and optional calibration multiplier for each selected
resource.  The admission step performs one all-or-none vector reservation:
either every nonzero component is added to its debt counter or no counter
changes.

Once admitted, an operation enters the same work-conserving dispatcher used
by every evaluated policy.  It then traverses only the stages implied by its
semantics.  A Write may consume HBM alone; an Atomic consumes an Atomic engine
and may also produce a reply; a Read consumes target memory followed by
response injection.  Each stage reports a local completion signal to the
resource table.  This organization separates three events that a conventional
single completion bit conflates: request acceptance, completion of a target
stage, and end-to-end operation completion.

SemaCredit operates after the endpoint has performed its ordinary opcode,
bounds, and address checks.  The demand estimate then chooses the admission
window for an already-valid operation.  An inaccurate estimate can hurt
performance by admitting a command too early or too late, but it does not
change the command's legality, decoded address, or operation semantics.

\subsection{Demand vector}

Let $\mathcal{R}$ be the set of receiver service resources.  In the current
design,

\begin{equation}
\mathcal{R}=\{\mathrm{HBM}_p\}\cup\{\mathrm{Atomic}_e\}
\cup\{\mathrm{Response}_j\}.
\end{equation}

An operation $i$ has demand vector $\mathbf{d}_i$, with zero entries for
resources it does not use.  The concrete estimator is

\begin{align}
d_{i,\mathrm{HBM}_p} &= b_i / R_p,\\
d_{i,\mathrm{Atomic}_e} &= \hat{t}_{\mathrm{atomic}}(op_i),\\
d_{i,\mathrm{Response}_j} &= \hat{t}_{0,j}+b_i^{resp}/R_j.
\end{align}

The implementation retains byte counters for HBM because streaming memory
traffic consumes more service as the transfer grows; $b_i/R_p$ is its
service-time view.  It uses explicit nanosecond debt for Atomic and response
injection because those stages include fixed per-operation work.  This
selective accounting was important in our experiments: charging every HBM
read a fixed service penalty unnecessarily throttled the MoE-shaped workload.

The target derives resource identifiers from fields that a remote-memory
endpoint already observes.  The opcode selects the operation class, address
decoding selects the HBM partition, response size selects the
response-injection demand, and endpoint configuration maps these demands to
local resource entries.  The resulting resource-scoped credits can be exposed
explicitly or carried by existing traffic classes or delivery contexts.
Endpoint validation and address decoding remain authoritative; the estimate
sets the admission timing for commands that have already passed those checks.

\subsection{Admission and phased return}

For each resource $r$, the receiver maintains outstanding debt $D_r$ and
window $W_r$.  Operation $i$ is admissible when

\begin{equation}
\forall r\;\mathrm{with}\;d_{i,r}>0:\quad D_r+d_{i,r}\le W_r.
\label{eq:admission}
\end{equation}

When an operation is admitted, SemaCredit reserves every resource named in its
demand vector in one all-or-none step.  This prevents partial reservations,
such as one command consuming the last HBM credit while another consumes the
last response credit and neither has a complete reservation.  Credit return
follows receiver execution rather than packet arrival.  HBM credit returns
after the memory access, Atomic credit after Atomic execution, and response
credit after the response is injected onto the return path.  Operation
completion is separate from credit return: a response-bearing operation
completes only after the response reaches the initiator, while a write without
a response completes after its target-memory stage.

This separation prevents two common accounting errors.  Returning all credit
at request arrival ignores target queues.  Holding all components until the
final response is safe but couples independent resources and reduces
pipeline parallelism.

Compared with returning the whole vector only at end-to-end completion, phased
return releases a resource as soon as that resource has finished its part of
the operation.  This can admit later operations earlier, because completed HBM
work no longer waits for response completion or Atomic execution.  It remains
safe because unfinished stages keep their reservations.  Section~\ref{sec:vector-necessity}
measures how much this earlier release helps in practice.

\subsection{Accounting invariants}

SemaCredit relies on three simple accounting invariants.  First, each resource
has its own outstanding debt $D_r$ and window $W_r$.  Admission increases
$D_r$ only after checking $D_r+d_{i,r}\le W_r$ for every resource named by the
operation, and stage completion only subtracts the frozen component that was
reserved at admission.  As a result, a resource's outstanding work stays
within its window as long as credit is returned after the corresponding stage
has actually executed.

Second, phased return is at least as permissive as end-to-end return under the
same arrivals, windows, and dispatcher.  Before a stage executes, both
policies keep its reservation.  After the stage executes, phased return frees
that component while end return continues to hold it.  Arrival-time return is
different: it can free response or Atomic debt before the receiver has done
the work, which is why it is treated only as an ablation.

Third, a single scalar window cannot replace the vector for arbitrary mixes.
Consider two receiver resources with safe window $W$ and two operations with
demands $(W,0)$ and $(0,W)$.  A scalar window of $W$ is safe but cannot admit
one of each operation concurrently, even though they use different resources.
A scalar window of $2W$ admits that pair, but it also admits two $(W,0)$
operations and overloads the first resource.  A scalar can still be retuned
for a known bottleneck; Section~\ref{sec:vector-necessity} tests how far that
retuning goes.

\subsection{Window sizing and arbitration}

A credit loop cannot keep a resource busy if its window covers less service
than the feedback delay.  SemaCredit therefore clamps the base window to the
measured worst-path RTT.  The simulated topology has a 1-$\mu$s worst-path
RTT, which becomes the HBM and response base window.  A serialized Atomic
engine uses four base windows to tolerate delivery variation caused by large
requests already queued in the forward fabric.  Section~\ref{sec:sensitivity}
varies both service latency and requested window; values below RTT are
automatically clamped.  A hardware implementation should tune the multiplier
from measured idle time and credit-turnaround delay rather than fix it
permanently.

Admission is work-conserving.  Unused capacity is not statically reserved for
small traffic.  Among currently admissible operations, the arbiter selects
latency-sensitive operations first, then shorter wire requests, then original
submission order.  This rule is applied to every evaluated policy, so gains do
not come from giving SemaCredit a uniquely favorable scheduler.

\subsection{Protocol mapping and state}

SemaCredit can use the receiver-grant or backpressure signal already provided
by the transport.  A minimal implementation keeps one debt counter and one
window counter per active receiver resource, plus the transport's existing
per-peer state.  With eight HBM partitions, one Atomic engine, and one response
engine, the simulated controller has ten resource entries.  This is comparable
to a baseline that tracks each receiver resource with a separate logical
admission context.  The important change is the meaning of the state: it tracks the
HBM, Atomic, and response work created by each admitted operation and returns
that work at the stage where it is consumed.

Ordering is orthogonal.  Writes to overlapping addresses, Atomics, fences, and
write-then-notify sequences retain the transport's ordering domain.  Credit
admission only decides when an operation may enter; the ordering domain still
decides which admitted operations may pass one another.

Admission can check all selected resources in parallel using one
add-and-compare lane per component, followed by an AND reduction and a
conditional counter write.  The command path therefore does not serialize ten
counter checks.  Arbitration occurs after admission and is outside the
reservation critical path.  The estimator can use table lookup for
Atomic opcodes and fixed-point multiply or shift-and-add for response size.

Table~\ref{tab:state} summarizes the logical state maintained by the
controller in the evaluated receiver configuration.  It assumes 32-bit debt
and window fields, a 48-bit service or feedback timestamp, and 16 bits of
multiplier and flags per resource.  A
command retains its frozen reservation until the corresponding stages return
it; rounding the resource mask, three 32-bit components, and phase flags to
16 bytes gives 2 KiB for the simulated 128 admitted commands.  Together with
ten resource entries, the added logical state is about 2.2 KiB.  A
hardware-oriented prototype can then replace this logical budget with SRAM,
area, frequency, and power measurements.

\begin{table}[!t]
\caption{Illustrative SemaCredit logical state budget.}
\label{tab:state}
\centering
\begin{tabular}{@{}p{0.22\columnwidth}p{0.46\columnwidth}rr@{}}
\toprule
Object & Added fields & Count & Bytes \\
\midrule
Resource entry & debt, window, timestamp, multiplier/flags & 10 & 160 \\
Command extension & mask, three frozen components, phase flags & 128 & 2,048 \\
\midrule
Total & excluding existing transport command state & & 2,208 \\
\bottomrule
\end{tabular}
\end{table}

\section{Methodology}

\subsection{Event model}

We use a deterministic Python event simulator that models four forward and
four return paths.  The simulator, plotting scripts, and configuration files
are prepared for artifact release.  We use multiple paths to introduce
mild timing heterogeneity while keeping routing orthogonal to the credit
policy.  Public scale-up fabric descriptions motivate the order of magnitude:
UALink 1.0 describes 200G-class lanes, a 100G option, credit-based flow
control, direct read/write/Atomic memory operations, and sub-microsecond
request-response latency over short accelerator links
\cite{ualink-whitepaper}; NVLink/NVSwitch systems similarly target
high-bandwidth, low-latency accelerator scale-up \cite{nvidia-nvlink}.  The
simulator therefore uses path rates of 200, 200, 100, and 200 Gb/s and
one-way propagation delays of 250, 300, 500, and 350 ns.  Requests select the
path with the earliest serialization completion.  A credit-delayed request
enters the network at its actual dispatch time, not its original application
submission time.

The target endpoint has eight HBM partitions at 512 Gb/s each, one 100-ns
serialized Atomic engine, and one 512-Gb/s response-injection engine with a
20-ns per-response overhead.  These defaults keep the bottlenecks
interpretable while matching the scale of accelerator endpoints: modern GPUs
expose multi-stack HBM systems with terabytes-per-second aggregate bandwidth
\cite{nvidia-hopper}, and public scale-up fabrics expose memory-style
Read/Write/Atomic operations over high-rate links \cite{ualink-whitepaper}.
The 100-ns Atomic point follows the upper range of published repeated-access
GPU Atomic measurements and is swept from 3 to 200 ns
\cite{gpu-atomic-contention}.  The response engine uses the same order of
bandwidth as a high-rate endpoint injection path, and the 20-ns fixed term is
swept from 0 to 40 ns to test sensitivity to per-response processing.  Reads
add a 40-ns target-read overhead.  Large write requests use 4,112 wire bytes;
read commands and other small requests use 80 bytes.  Small reads return 272
bytes, large reads return 4,112 bytes, and Atomics return 80 bytes.  A Read
consumes its returned payload length at HBM rather than only its
request-header length.
The default endpoint bound is 128 admitted target operations.  Each operation
progresses through its actual target stages and, when applicable, through the
return path before completion.  The sensitivity studies sweep credits, service
window, Atomic latency, response overhead, and mapping error, so the default
point anchors the stress setting while the sweeps show how conclusions change
around it.

\subsection{Parameterization and validation}

Table~\ref{tab:parameters} consolidates the default configuration.  The
forward and reverse fabrics are deliberately asymmetric so that path choice
and receiver service can disagree.  The 1-$\mu$s worst-path RTT is also useful
as a conservative base window: smaller requested values are clamped, whereas
larger values explicitly test excess buffering.  Timing perturbations change
serialization completion, application issue time, and bank choice while
leaving policy parameters fixed.  Every policy in a comparison receives the
same command trace and perturbation.

\begin{table}[!t]
\caption{Default simulator parameters.  Swept values are shown in parentheses.}
\label{tab:parameters}
\centering
\begin{tabular}{@{}p{0.39\columnwidth}p{0.54\columnwidth}@{}}
\toprule
Parameter & Setting \\
\midrule
Forward/reverse paths & 4 / 4 independent serialization queues \\
Path rates & 200, 200, 100, 200 Gb/s \\
One-way propagation & 250, 300, 500, 350 ns \\
Target HBM & 8 partitions, 512 Gb/s each \\
Atomic engine & one serialized engine, 100 ns (3--200 ns) \\
Response injection & 512 Gb/s plus 20 ns/response (0--40 ns) \\
Wire request size & 80 B small; 4,112 B large \\
Returned read size & 272 B small; 4,112 B large \\
Base service window & 1 $\mu$s (0.25--2 $\mu$s requested) \\
Endpoint command bound & 128 admitted commands (32--256) \\
Application experiments & five paired timing/workload perturbations \\
\bottomrule
\end{tabular}
\end{table}

We validate the simulator with 23 deterministic tests, which will be packaged
with the artifact.  The tests cover resource-window bounds, Atomic-chain
serialization, delayed dispatch without ``arrival in the past,'' completion of
every response-bearing policy, mapping error, reactive retries, release-time
ablations, and preservation of ordering domains.  Every reported run also
checks that all commands complete, proactive resource debt never becomes
negative, and the simulator terminates with no pending response event.  Oracle
and composed-policy equivalence tests cover cases where controllers should
coincide, helping catch accidental scheduler advantages.

The simulator is deterministic for a fixed trace.  We therefore use paired
perturbations rather than interpreting repeated deterministic runs as
independent statistical samples.  Main application bars report the mean of
five paired traces and the full minimum-to-maximum range.  The sensitivity and
robustness sweeps report all swept points, including settings where byte or
scalar accounting is competitive.

\subsection{Baselines}

We compare the following policies; the first six are the principal baselines:

\begin{itemize}
  \item \textbf{Aggregate byte}: one bounded destination pool, analogous to a
  destination/traffic-class receiver credit.
  \item \textbf{Aggregate+class}: aggregate control with a small-operation
  class reserve.
  \item \textbf{Resource byte}: a strong composition with a byte window per
  HBM, Atomic, and response resource and the same work-conserving priority.
  \item \textbf{Primary service}: service-time accounting for HBM and Atomic
  resources at the primary target, but no response-injection reservation.
  \item \textbf{Scalar service}: an idealized normalized-operation controller
  that sums all predicted endpoint stages into one scalar cost.  We sweep its
  window across the explored range.
  \item \textbf{Reactive exhaustion}: a constructed policy that uses
  UET-style resource-exhaustion and recovery primitives to disable an
  exhausted semantic resource and retry after feedback.
  \item \textbf{SemaCredit}: selective HBM byte occupancy plus Atomic and
  response service-time components.
  \item \textbf{Arrival/end return}: SemaCredit ablations that return the full
  vector at target arrival or hold it until end-to-end completion.
  \item \textbf{Adaptive SemaCredit}: SemaCredit with an optional
  exponentially weighted moving average (EWMA) multiplier for systematic
  prediction error \cite{jacobson88}.
  \item \textbf{Oracle}: SemaCredit with perfect address-to-resource mapping.
\end{itemize}

Resource byte gives each modeled target resource its own byte window, making
it a stringent composition baseline rather than a simple aggregate-credit
controller.  Primary service is the principal ablation for the downstream
response component.  Inspired by Harmonic's normalized operation-cost
accounting \cite{harmonic}, scalar service gives the controller
operation-level visibility into every modeled endpoint stage but compresses
the stages into one cost.

\subsection{Workloads and metrics}

The three microbenchmarks isolate HBM hotspot, shared Atomic-engine contention,
and response incast.  Each includes periodic latency-sensitive operations.  We
also construct three controlled \emph{application-shaped} mixes derived from
common accelerator communication phases.  The AllReduce-shaped mix represents
bulk collective chunks \cite{nccl,msccl,taccl} interleaved with small
reduction or completion Atomics that decide when a phase can advance.  The MoE-shaped mix
represents skewed expert traffic and routing metadata, as in sparse expert
models \cite{gshard,switch-transformers,deepspeed-moe}.  The
remote-read-shaped mix represents parameter, KV, or metadata fetches, where
small requests can generate large or frequent responses.  The three mixes
provide controlled coverage of Atomic-heavy, HBM-skew-heavy, and
response-heavy receiver pressure.  Each application
experiment uses five seeds that perturb timing and, where applicable, expert
or partition choice.

The primary metric is P99 latency of the small operations.  We also report
bulk P99, makespan-derived useful goodput, maximum per-resource queueing delay,
maximum target operations in flight, and controller state entries.  CSV inputs,
simulation code, plotting scripts, and tests accompany the paper.

\section{Evaluation}

\subsection{Does semantic accounting help?}

\begin{figure}[!t]
  \centering
  \includegraphics[width=\columnwidth]{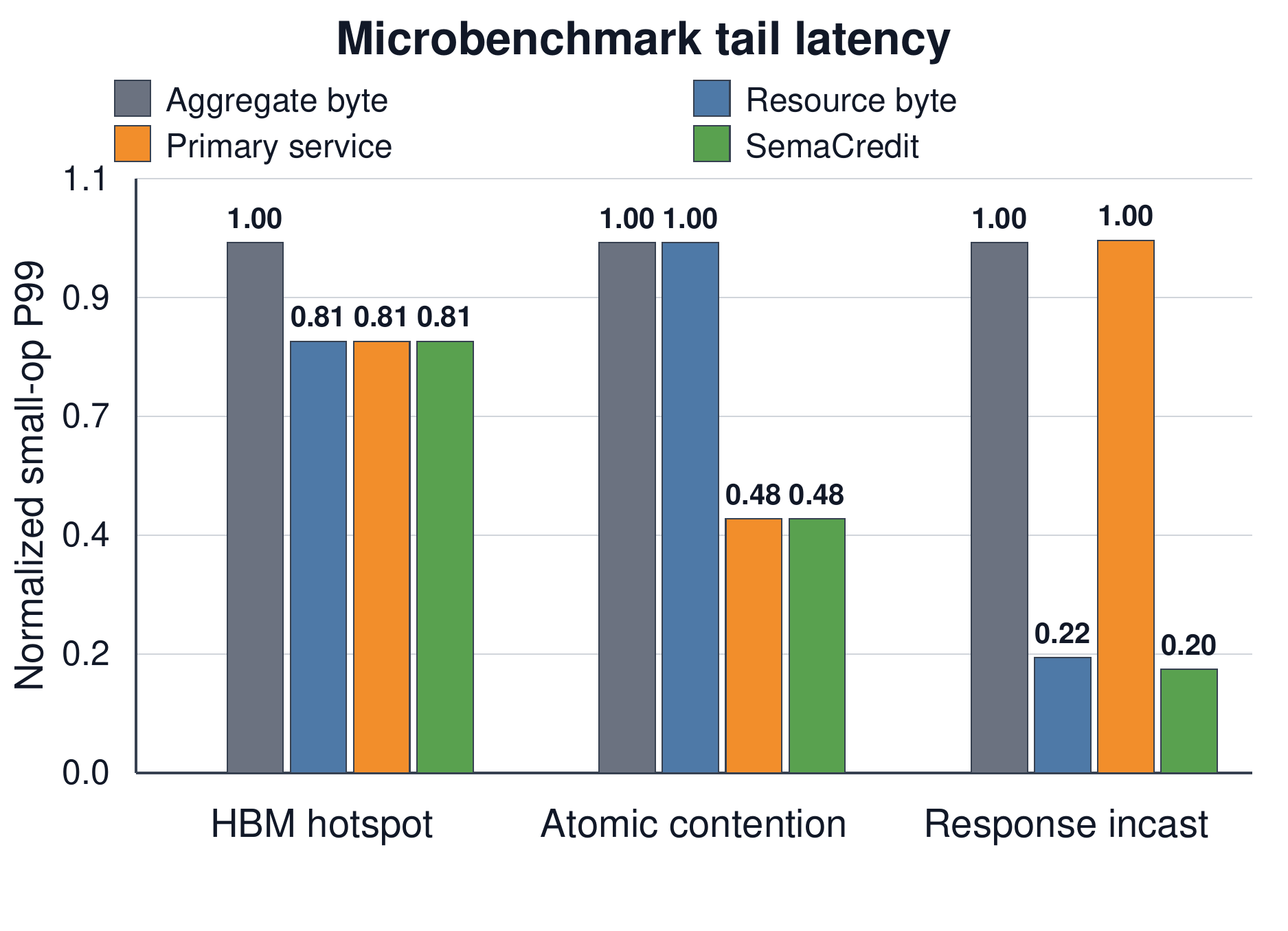}
  \caption[Small-operation P99 in the three resource-isolation
  microbenchmarks]{Small-operation P99 in the three resource-isolation
  microbenchmarks, normalized to aggregate byte credit.  Resource byte is
  sufficient for HBM, primary service is sufficient for Atomic contention,
  and the full vector is required for response incast.}
  \label{fig:micro}
\end{figure}

Figure~\ref{fig:micro} normalizes small-operation P99 to aggregate byte credit.
HBM hotspot is the control case: resource byte, primary service, and SemaCredit
all reach approximately 0.81$\times$ aggregate latency.  SemaCredit is within
0.1\% of resource byte in latency while delivering 3.5\% higher goodput across
the five timing perturbations.  This confirms that a byte controller is already
strong when the receiver work mostly grows with the number of bytes.

Under Atomic contention, aggregate and per-resource byte control are both
1.0$\times$: separating the Atomic resource provides no benefit if its small requests
are still budgeted as memory bytes.  Primary service and SemaCredit reduce P99
to 0.48$\times$, a 52.4\% mean improvement over resource byte across five
timing perturbations, with identical goodput.  Maximum Atomic queueing falls
because the controller bounds predicted serialized work rather than packet
bytes.

Response incast distinguishes primary service from the full vector.  Primary
service remains at aggregate performance because it releases the operation
from target control before accounting for response injection.  Resource byte
reduces P99 to 0.22$\times$ aggregate by separating the response resource.
SemaCredit further reduces it to 0.20$\times$: its mean improvement over the
strong resource-byte baseline is 10.2\%, with 0.2\% higher goodput.  The
incremental benefit comes from the response engine's per-response overhead;
when that overhead is zero, byte and service accounting converge.

\subsection{Is a completion-path vector necessary?}\label{sec:vector-necessity}

\begin{figure}[!t]
  \centering
  \includegraphics[width=\columnwidth]{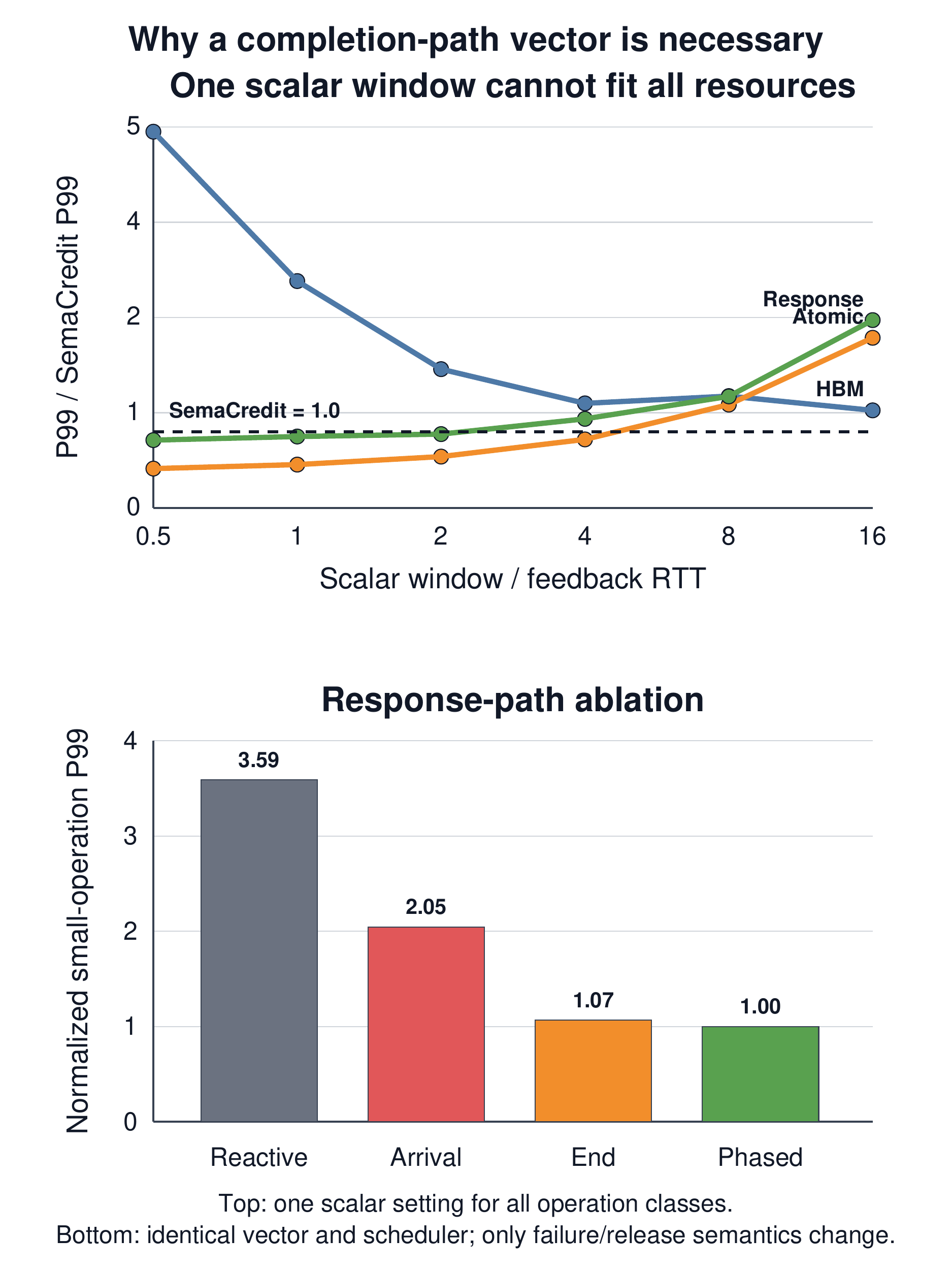}
  \caption{Top: a single service window is swept across the same workloads;
  settings that help one bottleneck waste parallelism or overload another.
  Bottom: response-incast P99 under reactive exhaustion and vector-return
  ablations, normalized to phased SemaCredit.}
  \label{fig:strong}
\end{figure}

Figure~\ref{fig:strong} tests two stronger alternatives.  Motivated by
Harmonic's normalized operation-cost accounting \cite{harmonic}, scalar
service sums every predicted stage into one operation cost and sweeps its
window from 0.25 to 16 feedback RTTs.  Retuning
the scalar can improve one workload, but no fixed setting matches the vector.
At a scale of four, scalar and SemaCredit have equal goodput in all three
microbenchmarks; scalar P99 is 10\% lower for Atomic traffic, but 37\% higher
for HBM and 17\% higher for response traffic.  At a scale of two it improves
Atomic and response P99, but HBM P99 is 1.82$\times$ SemaCredit and response
goodput is 3.1\% lower.  The scalar must guess the active bottleneck; the vector
names it.

Reactive semantic exhaustion waits for a queue to cross its window, disables
the resource, and retries after feedback and recovery using UET-style
resource-exhaustion primitives.  Relative to this baseline, SemaCredit
reduces Atomic and response P99 by 50.1\% and 72.1\%, avoids 92 and 34 mean
retries, and improves goodput by 4.5\% and 3.9\%.

Finally, returning all debt at target arrival makes response P99
2.05$\times$ phased return because downstream work is no longer bounded.
Holding the full vector until end-to-end completion is safe but has 6.4\%
higher P99 and 0.9\% lower goodput because completed HBM work remains coupled
to response completion.  Atomic-only traffic leaves the release variants tied,
a useful check that phased return matters when an operation spans multiple
receiver stages.  Section~\ref{sec:robustness} separately examines whether a
normalized multiplier can protect against estimator error.

\subsection{Application-shaped mixes}

\begin{figure}[!t]
  \centering
  \includegraphics[width=\columnwidth]{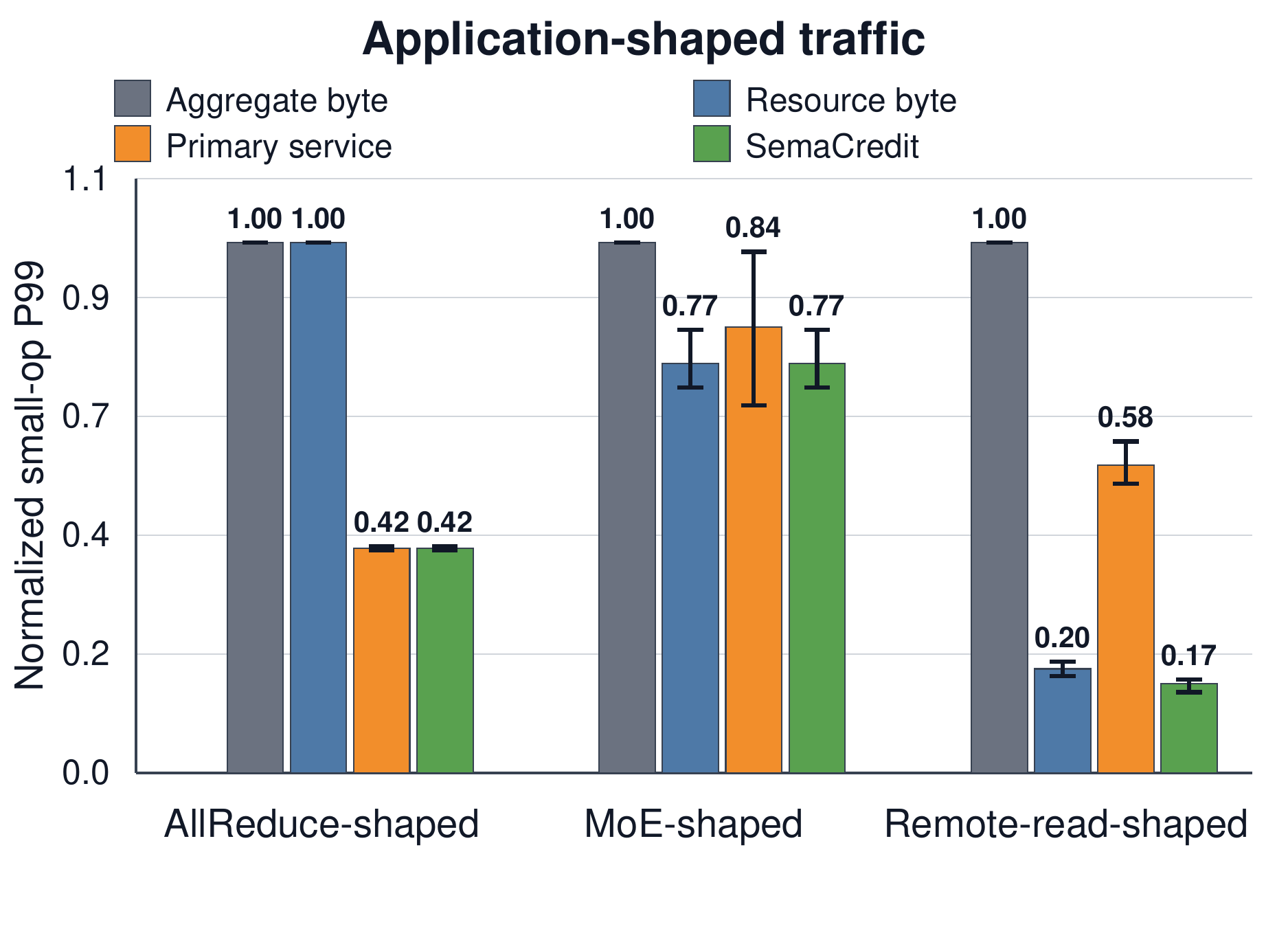}
  \caption[Application-shaped traffic]{Mean normalized
  small-operation P99 for application-shaped traffic.
  AllReduce-shaped combines chunk writes with reduction/completion Atomics;
  MoE-shaped combines skewed expert writes with routing-metadata reads;
  remote-read-shaped combines large parameter/KV reads with small control
  reads.  Error bars span five timing/workload perturbations.}
  \label{fig:applications}
\end{figure}

Figure~\ref{fig:applications} reports the five-seed application-shaped results.
For AllReduce-shaped traffic, primary service and SemaCredit reduce P99 by
57.7\% relative to both aggregate and resource byte.  This mix stresses the
serialized Atomic engine.  SemaCredit's useful goodput is 1.9\% below resource
byte; the small loss points to room for window autotuning.

The MoE-shaped workload is dominated by skewed HBM writes.  Resource byte and
SemaCredit are tied and reduce P99 by 22.8\% versus aggregate; goodput is tied.
This is a useful control case: after a strong resource-byte controller protects
the hot HBM partitions, operation semantics add little extra benefit.

Remote-read-shaped traffic stresses response injection.  SemaCredit reduces
P99 by 83.3\% versus aggregate, 14.5\% versus resource byte, and 71.1\% versus
primary service.  Its goodput is within 0.1\% of resource byte.  The large gap
from primary service shows why an operation must reserve downstream resources,
not only the resource named by its target address.

\subsection{Sensitivity}\label{sec:sensitivity}

The sensitivity sweep varies the two costs that do not simply scale with
request bytes.  At Atomic latencies of 3, 10, 30, 100, and 200 ns,
SemaCredit's P99 benefit over resource byte is 50.3\%, 47.2\%, 47.4\%,
57.8\%, and 72.5\%.  At 3--30 ns, primary
service provides little benefit but the full vector helps because response injection has
become the dominant downstream stage.  At 100--200 ns, primary service and
SemaCredit tie because the Atomic stage dominates.  This crossover, rather
than Atomic cost alone, motivates completion-path admission.  As response
fixed overhead rises through 0, 10, 20, and 40 ns, the corresponding benefit
is 0\%, $-1.2\%$, 12.9\%, and 24.2\%; the 10-ns point is slightly negative,
showing that service accounting becomes most useful once fixed response cost
is non-negligible.

With 32 total credits, the Atomic microbenchmark is already tightly bounded and
SemaCredit ties resource byte.  Its Atomic P99 benefit grows to 26.9\%, 57.8\%,
and 61.5\% at 64, 128, and 256 credits.  Response-incast improvement remains
15.3\% across those four settings in the smaller sensitivity workload.  A
requested service window below the 1-$\mu$s feedback RTT is clamped and gives
identical results at 250, 500, and 1,000 ns.  Expanding the base window to
2,000 ns weakens Atomic isolation, as expected, without reducing goodput.

Finally, address-to-HBM mapping error degrades resource-aware control.  At
0\%, 10\%, 25\%, and 50\% injected error, SemaCredit improves P99 over
aggregate by 26.9\%, 23.9\%, 18.8\%, and 17.0\%, respectively.  At 25\% error
it is 5.6\% slower than the per-resource byte baseline, showing that partition
prediction quality matters for performance.  Atomic and response resources are
opcode-derived in the current model, so this sweep isolates address-to-HBM
classification.

\subsection{Robustness to model error}\label{sec:robustness}

\begin{figure}[!b]
  \centering
  \includegraphics[width=\columnwidth]{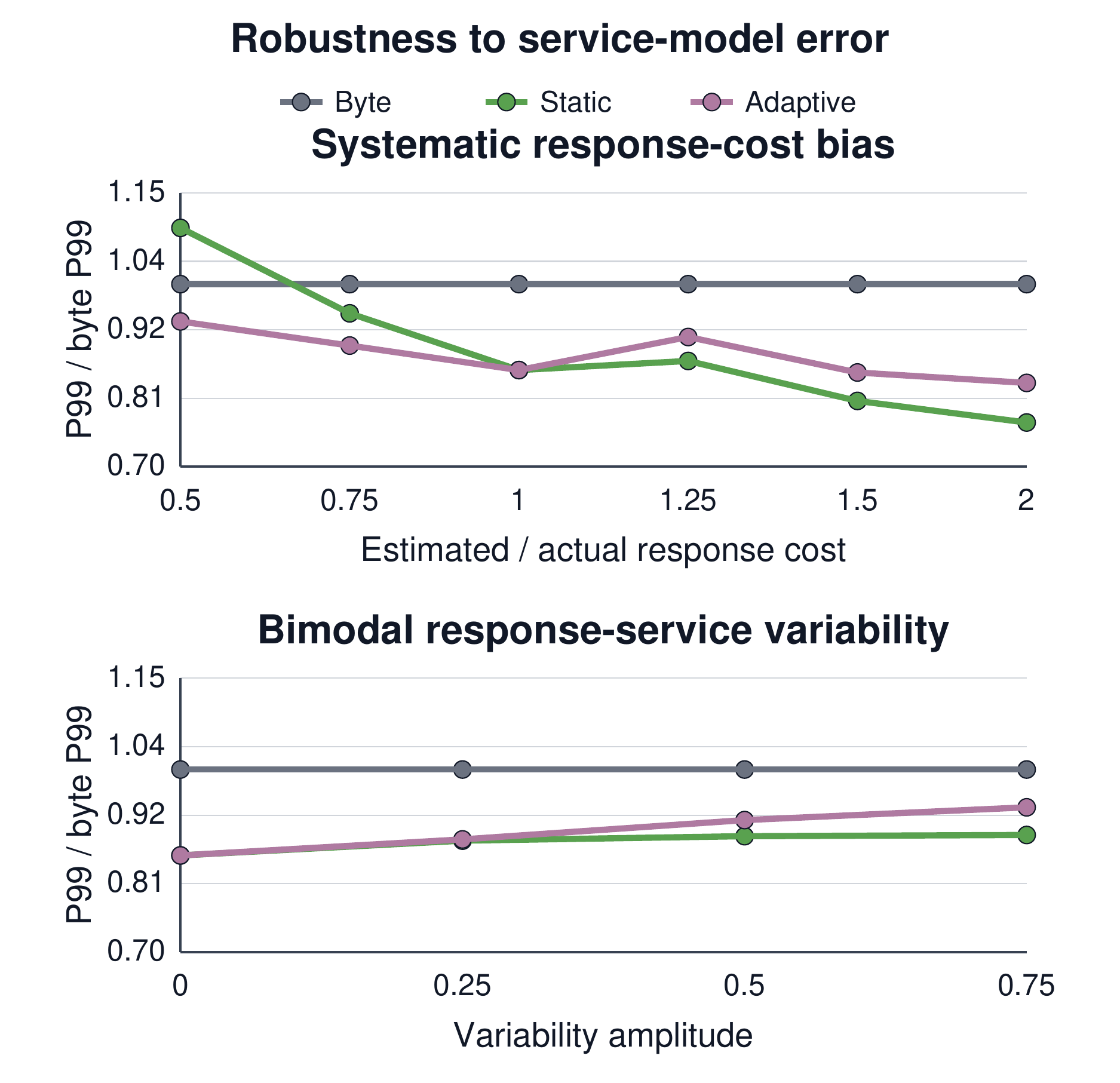}
  \caption{Response-incast robustness, normalized to the per-resource byte
  baseline.  Top: systematic cost-estimation bias.  Bottom: stationary
  bimodal service variability.  Lower P99 is better; the byte baseline is
  normalized independently at every point.}
  \label{fig:robustness}
\end{figure}

This section asks how much SemaCredit depends on accurate cost estimates.  The
vector identifies which resource is consumed, but fixed-cost entries such as
response overhead and Atomic service still require calibration.
Figure~\ref{fig:robustness} separates systematic bias from run-to-run service
variation.  The static policy freezes the demand recorded at admission.  The
adaptive policy additionally maintains one multiplier per service resource and
updates it at completion using an EWMA, a standard smoothing method in network
estimators \cite{jacobson88}:
$m\leftarrow0.9375m+0.0625(t_{obs}/\hat{t})$.  Existing command reservations
remain frozen, preventing an estimator update from returning a different
amount than was admitted.

With a 50\% systematic underestimate, static SemaCredit is 9.2\% worse than
resource byte because the response pipeline is admitted too aggressively.
Adaptive SemaCredit instead becomes 6.1\% better with the same goodput.  At
the correct estimate both variants are 14.1\% better and deliver 0.3\% higher
goodput.  A twofold overestimate reduces static P99 by 22.7\%, but the lower
latency comes from throttling: goodput falls by 8.2\%.  Adaptation
limits the goodput loss to 2.0\% while retaining a 16.2\% P99 reduction.

The lower panel separates persistent model bias from random per-operation
variation.  Under zero-mean bimodal
variability, the static mean-cost controller has lower P99 than the adaptive
variant at the largest amplitude: 10.8\% versus 6.2\% below resource byte,
with both goodputs within 0.6\% of the byte baseline.  The EWMA reacts to
individual slow samples even though the long-run mean is unchanged.  The
lesson is to use slow calibration for persistent bias and avoid chasing every
random sample with per-operation cost updates.  A deployment should separate
window autotuning, which responds to persistent idle time, from per-operation
cost calibration.

\section{Discussion and Limitations}

\subsection{Deployment and calibration}

SemaCredit can expose resource vectors explicitly or derive them inside the
receiver.  In an explicit interface, the receiver advertises opaque
resource-class credits and the initiator attaches the selected class mask to
each command; the receiver treats that mask as an admission hint and computes
the final resource entries from its own validation and address decoding.  In
the implicit interface used in our evaluation, the wire protocol is unchanged:
the target classifier maps existing opcode, address, length, and response-size
fields to the local resource table before admission.  This avoids exposing
physical partition identifiers while still giving the receiver resource-scoped
control.

Feedback need not expose a counter update for every stage completion.  The
receiver may coalesce newly available debt into the transport's next grant or
backpressure message, just as byte credits are commonly aggregated.  Local
admission uses the precise counter; delayed feedback only postpones remote
dispatch and cannot oversubscribe the target.  A practical implementation can
also quantize nanosecond debt into small service quanta.  Quantization rounds
demand upward for protection; evaluating its utilization loss is left to a
hardware-oriented follow-up.

Calibration uses endpoint observations rather than application annotations.
HBM demand follows bytes and measured partition rate.  Atomic lookup entries
can be initialized per opcode, while response cost combines a fixed injection
term with returned bytes.  Busy/idle cycles and completion timestamps provide
the signals needed to update a slow multiplier.  The receiver owns these
measurements and may cap multipliers and windows, preventing a sender from
claiming an artificially small demand.  Rapid workload phase changes remain a
risk: an estimator that reacts too slowly under-admits, while one that reacts
to every sample reproduces the stationary-variability penalty in
Figure~\ref{fig:robustness}.

\textbf{Relationship to existing receiver controls.}  Existing transports
already provide receiver-driven grants, byte credits, host backpressure,
per-resource contexts, and normalized verb-cost accounting
\cite{homa,sird,meta-roce,harmonic}.  SemaCredit builds on these ideas at a
different granularity: each remote-memory operation reserves the endpoint
stages it will use---such as HBM, Atomic execution, and response
injection---and returns each reservation when that stage completes.  The
scalar-service baseline tests whether preserving this vector and its phased
lifetime matters in practice.

\textbf{Information source.}  Opcode and response length are explicit.
Address-to-HBM mapping may be known to the endpoint but hidden from the
initiator.  A deployment can therefore let the target issue resource-scoped
credits without exposing physical partition details.  Coarse bank groups or
learned conflict classes may be sufficient.  Our error experiment suggests
that the best classifier granularity depends on prediction noise.

\textbf{Window adaptation.}  The current prototype clamps to feedback RTT and
uses a four-RTT Atomic window.  A hardware design should derive these windows
from measured idle time, feedback delay, and target queue occupancy.  The
optional service estimator tracks a normalized multiplier with an EWMA and
recovers systematic response-cost underestimation, but static mean costs are
better for stationary heavy-tailed service.  Window adaptation should be
studied separately from cost calibration and compared against a similarly
adaptive byte baseline.

\textbf{Reliability and ordering.}  Integrating SemaCredit with reliability
requires preserving one semantic reservation across retransmission and
duplicate filtering.  A retransmitted command should reuse the original
reservation, and a duplicate completion should leave credit unchanged.  These
rules can use the transport's operation identifier.  Ordering domains for
overlapping writes, Atomics, fences, and notifications compose naturally with
credit admission because admission only delays when an operation may enter.

\textbf{Evaluation scope and next validation step.}  The deterministic model
isolates receiver completion-path accounting as a first step toward hardware
validation.  It captures operation mixes, endpoint-stage
contention, multipath timing, and resource-specific credit lifetimes, while
leaving framework scheduling, kernel execution, and product-specific
microarchitecture to the next validation stage.  The natural follow-up is to
replay collected accelerator traces in a packet/network simulator and then
validate the controller in an RTL or FPGA receiver model.

\section{Future Work}

The present simulator establishes a mechanism-level result: independently
queued receiver stages are better represented as distinct completion-path
resources than as one byte or service counter.  Turning that result into a
deployable scale-up transport requires four follow-on steps.

\textbf{Measured traces and resource calibration.}  We first plan to
instrument GPU and AI-accelerator communication paths for collective
training, MoE dispatch, parameter/KV reads, and fine-grained synchronization.
A trace will record operation type, request and response length, destination,
dependency domain, submission phase, and receiver-stage timestamps.  These
measurements will replace the current constant HBM, Atomic, and response costs
with empirical distributions and burst structure.  We will retain the
synthetic microbenchmarks as controlled diagnoses, but replay collected traces
for application claims.  The first criterion is predictive fidelity: whether
the demand vector forecasts the identity and depth of the measured bottleneck.

\textbf{Adaptive control under delayed feedback.}  Capacity adaptation and
per-operation cost estimation should be separated.  We will study target-owned
estimators under phase changes, heavy-tailed service, coalesced grants, and
multiple initiators, and derive conditions that bound debt despite feedback
delay and quantization.  Comparisons will include equally adaptive byte and
scalar-service controllers as well as static baselines.  Beyond latency and
goodput, this stage will measure utilization loss, fairness, convergence time,
and sensitivity to classification error.

\textbf{Protocol and hardware realization.}  We next plan a cycle-accurate
implementation of classification, atomic vector reservation, staged credit
return, and grant coalescing, followed by an RTL/FPGA prototype when the
interface is stable.  Mapping the controller onto UET/MRC-style resource and
feedback primitives will determine which changes are endpoint-local and which
require wire-visible resource classes.  Synthesis and prototype measurements
will replace the logical state estimate with SRAM, area, frequency, power, and
feedback-bandwidth costs, including the effects of counter quantization.

\textbf{Reliability and end-to-end validation.}  The controller must reserve
semantic work once across loss and retransmission, reject duplicate returns,
and compose with fences and Atomic ordering without deadlock.  We will extend
the model to failures, multiple receivers, and competing tenants before
integrating the non-coherent interface with collective and inference runtimes.
End-to-end evaluation will report training-step or token latency, communication
overlap, accelerator utilization, and energy, connecting transport-level P99
to application-visible outcomes.  These studies will clarify the traffic
regimes where completion-path credits provide durable benefit.

\section{Related Work}

Scale-up and adjacent memory fabrics expose the context in which SemaCredit
would operate.  UALink targets accelerator scale-up with direct
Read/Write/Atomic memory operations, credit-based flow control, and 100G/200G
link options \cite{ualink-whitepaper}.  NVLink/NVSwitch systems show the same
pressure toward high-bandwidth, low-latency GPU fabrics
\cite{nvidia-nvlink}, while CXL and UCIe illustrate adjacent memory and
chiplet fabrics with their own flow-control and transaction abstractions
\cite{cxl-spec,ucie-spec}.

Recent transport proposals provide the grant and backpressure mechanisms that
could carry resource-aware admission.  UET specifies receiver congestion
control, packet-delivery contexts, remote memory operations, reliability, and
resource-exhaustion recovery \cite{uetspec}.  SemaCredit supplies a predictive
admission policy for the endpoint stages behind such mechanisms.  MRC extends
RoCEv2 with multipath, bounded flight, explicit host backpressure, and
composable controller primitives \cite{mrc}.

Meta's production design combines receiver-driven CTS admission with
collective channel buffers and GPU-aware scheduling \cite{meta-roce}.  It is
the closest evidence that receiver memory readiness should influence AI
transport.  Our work generalizes the control object from collective channels
to Read/Write/Atomic operation demand vectors.  SIRD \cite{sird}, Homa
\cite{homa}, ExpressPass \cite{expresspass}, and NDP \cite{ndp} demonstrate
the value of receiver-driven grants, scheduled traffic, or receiver-link
scheduling.  SemaCredit focuses on what the receiver should grant when
endpoint service resources, rather than only the downlink, are contended.

Harmonic monitors BPS, normalized verb-processing rate, PCIe bandwidth, and
cache interference, then enforces per-tenant RDMA performance isolation via a
rate-control loop \cite{harmonic}.  DCQCN, TIMELY, and HPCC address
large-scale RDMA or datacenter congestion control from the network-feedback
side \cite{dcqcn,timely,hpcc}.  SemaCredit targets a different control point:
sub-operation admission at the receiver, where one command can simultaneously
reserve a target partition and a downstream response engine and release those
components separately.  We compare with an idealized scalar-service
controller that gives the scalar approach direct operation-level visibility.

RDCA measures receiver memory contention in production and bypasses DRAM with
a cache-resident receiver service \cite{rdca}.  SRNIC reduces RNIC connection
state and supports many performant RDMA connections \cite{srnic}.  These works
motivate endpoint bottlenecks and scalable state, respectively.  SemaCredit
addresses the complementary admission question: which semantic work should
enter the existing endpoint pipelines, and when should each reserved stage be
released?

Multipath RDMA mechanisms such as MP-RDMA address path utilization and
reordering \cite{mprdma}.  We use multipath serialization queues in the model,
but path selection is held constant across policies.

Collective and sparse-model systems motivate the application-shaped mixes.
NCCL, MSCCL, TACCL, and SwitchML optimize collective communication and
aggregation for distributed training \cite{nccl,msccl,taccl,switchml}.
GShard, Switch Transformers, and DeepSpeed-MoE show how sparse expert models
create skewed expert dispatch and metadata traffic
\cite{gshard,switch-transformers,deepspeed-moe}.  SemaCredit uses controlled
mixes inspired by these phases to isolate receiver resources.

\section{Conclusion}

Byte credits work well when receiver work mainly follows transfer size.
SemaCredit extends that accounting to operations whose completion path also
uses Atomic execution or response injection.  It associates each remote-memory
operation with a vector of HBM, Atomic, and response demands and returns each
reservation at the stage that consumes it.  Simulation results show that
per-resource bytes, one normalized total-service scalar, and reactive resource
exhaustion each miss part of this behavior across HBM, Atomic, and response
pressure.  The strongest case appears when small operations gate progress
through Atomic or response resources; the next step is to replay measured
traces and test the controller in a hardware-realistic receiver.

\bibliographystyle{ACM-Reference-Format}
\bibliography{references}

\end{document}